# The origin and fate of $O_2$ in Europa's ice: an atmospheric perspective

R.E. Johnson[1,2], A.V. Oza[3], F. Leblanc[4], C. Schmidt[5], T.A. Nordheim[6]

[1] University of Virginia, Charlottesville, VA 22904 USA &
[2] Physics Department, NYU, New York, NY 10003
[3] Physikalisches Institut, Universität Bern, Gesellschaftsstrasse 6,
   CH-3012 Bern, Switzerland
[4] LATMOS/IPSL, UPMC Univ. Paris 06 Sorbonne Universités, UVSQ, CNRS,
   4 place Jussieu, Paris 75005, France
[5] Center for Space Physics, Boston University, Boston MA 02215 USA
[6] Jet Propulsion Laboratory, California Institute of Technology, Pasadena, California, U.S.A.,

**Abstract**

The early prediction and subsequent detection of an $O_2$ atmosphere on Europa, coupled with the discovery that Europa has an ocean under its ice mantle, has made this moon a prime astrobiologic target, soon to be visited by the JUICE and Europa Clipper spacecraft. In spite of the considerable number of observational, modeling, and laboratory efforts, understanding the physics leading to the observed morphology of Europa's near surface $O_2$ atmosphere has been problematic. This is the case as the observed emissions depend on the local incident plasma ion flux, the local temperature and composition of the regolith, as well as on the near surface electron temperature and density. Here we rely heavily on earlier reviews briefly summarizing the observational, laboratory and simulation efforts. Although it is agreed that radiolysis of the surface ice by the incident Jovian plasma is the ultimate source of observed $O_2$, a recent, simple model of thermal desorption from a regolith permeated with $O_2$ has changed the usual paradigm. This suggests that the observed orbital dependence of the local source of the near-surface $O_2$ atmosphere is due to thermal release of $O_2$ likely trapped on the ice grains at dangling bonds with a smaller contribution due to direct sputtering. This could also impact our understanding of the suggestion that the radiolytic products in Europa's regolith might be a source of oxidants for its underground ocean.

**Introduction**

Based on laboratory data, Jupiter's moon Europa was predicted to have a tenuous $O_2$ atmosphere due to the radiolytic decomposition of its icy surface by the incident Jovian plasma particles (Johnson et al. 1982; Johnson 1990). Although this prediction has been borne out by extensive Hubble Space Telescope (HST) observations (e.g., Hall et al. 1995; 1998; McGrath et al. 2004; Roth et al. 2016) and by additional laboratory data (e.g., see summaries in Johnson 2011; Teolis et al. 2017a), a large number of uncertainties remain as to how the laboratory data should be applied at Europa (e.g., see Plainaki et al. 2018). That is, it is agreed upon that $O_2$, as well as its concomitant radiolytic product $H_2$, are produced and ejected by the incident plasma, but a clear, quantitative description of the spatial and temporal distribution of the oxygen aurora based on HST observations remains elusive. This is unfortunate as Europa is a prioritized exploration target for both NASA and ESA and will soon be visited by spacecraft with the principal goal of understanding its potential as a habitable world. Therefore, there has been a significant recent increase in the observational and modeling studies of the ambient gas-phase $O_2$ at Europa.

It is now established that incident plasma particles and short wavelength solar photons can decompose ice and produce $O_2$ and $H_2$. This is the case not only at Europa, but also at its neighbors Ganymede (e.g., Spencer et al. 1995; Calvin et al. 1996; Turc et al. 2014; Leblanc et al. 2017) and Callisto (Spencer and Calvin 2002; Vorburger et al. 2015), as well as at Saturn's rings (Johnson et al. 2006; Tseng et al. 2010) and its small icy satellites (e.g., Tokar et al. 2012; Teolis et al. 2016). The production of oxygen in Europa's surface ice is especially interesting, as it has been suggested to be important for potential biologic activity by subduction of oxygen rich material into its underground ocean (e.g., Chyba 2000; Johnson et al. 2003; Orlando et al. 2005; Greenberg 2008; Vance et al. 2016; Russel et al. 2017). This possibility depends critically on the state of the $O_2$ produced in Europa's icy regolith: that is, on how $O_2$ binds in and diffuses through the porous icy regolith and ice mantle. It also depends on the coupling between Europa's surface ice and its putative ocean. That there is coupling has been discussed for years based on the modeling of the images of its nearly craterless and fractured surface (e.g., Pappalardo et al. 2009; Parro et al. 2016; Kattenhorn & Prockter 2014; Prockter et al. 2017). Recent ground-based observations of Europa point towards the presence of endogenic materials (possibly chlorinated salts) on the surface (Brown and Hand, 2013; Fischer et al., 2015, 2016; Ligier et al., 2016), and irradiation experiments by Hand & Carlson (2015) suggest that NaCl-rich material from the sub-surface ocean may explain Europa's yellow-brownish color at visible wavelengths. In addition, the ambient sodium and potassium emissions observed (Brown et al. 1996; Brown 2001) also suggest that there is coupling between the surface and the likely 'salty' ocean (e.g., Johnson 2000; Leblanc et al. 2002; 2005). Added to these observations, the analysis of Galileo magnetometer measurements has revealed that Europa is a source of $Na^+$, $Cl^-$ and $Cl^+$, and $K^+$ pickup ions to the Jovian magnetosphere (Desai et al., 2017; Volwerk et al., 2010, 2001). Finally, recent observations indicating possible plume activity on Europa (e.g., Roth et al. 2014; Sparks et al. 2016; 2017) has significantly increased interest in the importance of the coupling between its surface and ocean.

Although there is now considerable laboratory data on the sputtering of ice, the surprising inability to carry out a definitive simulation of the observed oxygen aurora is due to the complexity of processes influencing emission. Emission depends on the local plasma ion temperature, composition, precipitation rate and sputtering yield, as well as on the local surface composition affecting the production rate, and local electron temperature and density responsible for collisional excitations. Moreover, it is due in part due to our lack of understanding of the interactions of $O_2$ in the porous regolith (e.g., Johnson et al. 2005; Loeffler et al. 2006; Laufer et al. 2017) and our understanding on how the composition and nature of that regolith affect those interactions as discussed below.

There are numerous and even very recent reviews on Europa's surface properties, the plasma-induced production of $O_2$, and on the observational database (e.g., Carlson et al. 2009; Johnson et al. 2009; 2013; McGrath et al. 2004; Cassidy et al. 2010; Gudipati & Castillo-Rogez 2013; Plainaki et al. 2018). Rather than a new review, we will refer to these in the following focusing on how the modeling of the *gas-phase observational data* affects our understanding of the likely state of $O_2$ in Europa's regolith. We first give a brief review on the production of $O_2$, then summarize the observational data, and, finally, discuss our recent understanding.

**Radiolysis In Water Ice**

As often described, radiation absorbed by gas phase water molecules can directly break bonds, forming $H + OH$ with a smaller fraction of $H_2 + O$. It can also cause ionization, with the ionized species eventually recombining with an electron also producing fragment radicals. This is also the case for electron impact excitation producing excited H, which was the basis for the first suggestion of water vapor plumes on Europa (Roth et al. 2014). When $H_2O$ molecules in water ice absorb radiation, the results are more complicated due to trapping (e.g., Johnson et al. 2005) and the rapid reaction of an excited or ionized species with its neighbors. These interactions can depend on the energy density absorbed locally, as closely spaced ionized or excited species interact efficiently (e.g., Johnson 2011). Ignoring the details of these interactions, the net result of irradiation experiments are often given in terms of product yields called G-values: the number of a particular product produced per 100eV of energy absorption. In addition to the radicals, relatively stable new molecules are produced: $H_2$, $O_2$ and $H_2O_2$ (e.g., Carlson et al. 2009).

The surprisingly large yield of gas phase $O_2$ emanating from irradiated ice seen in early experiments was suggested to be due to the presence of the vacuum interface at the sample's surface resulting in the loss of the most volatile product, $H_2$ (e.g., Brown et al. 1982). That is, prompt loss of $H_2$, even at temperatures well below that of Europa's surface, leaves an oxidized composition enhancing the efficiency for eventually producing $O_2$ by the subsequently absorbed radiation or, if irradiation occurs at very low temperature, by increasing the temperature. The former is seen directly in laboratory experiments in as terms of the yield vs. irradiation time at low fluences (e.g., Brown et al. 1982; Johnson et al. 1983) and has been confirmed in subsequent work (e.g., Teolis et al. 2017). Depending on the surface temperature, $O_2$ is, of course, also volatile. Therefore, it can escape by thermal desorption, be driven from the surface by the radiation flux, and can diffuse into the regolith, as we will discuss shortly.

Although the exact details for the formation of $O_2$ in ice is still uncertain and the database is incomplete, there is now a considerable body of laboratory data on the net yield of $O_2$ (number of $O_2$ per incident ion) as a function of both the energy density deposited in the icy surface and the surface temperature (e.g., Teolis et al. 2017 and references herein). There are also measurements of the amount of $H_2O_2$ in the ice following a given radiation dose (e.g., Carlson et al. 1999a; Carlson et al. 2009; Hand & Carlson 2011). However, one major caveat for application to planetary science is the very limited data on the role of trace contaminants on enhancing or quenching the production of $O_2$, although recent observations at the icy Saturnian satellites suggest that it is very important (Teolis et al. 2016). The irradiation of an icy surface also produces damage in the regolith ice grains (interstitials, vacancies, defects and incomplete bonds) which has been suggested as being detected at Europa as an crystalline to amorphous transition at depths < 1mm (Hansen & McCord 2004; Paranicas et al. 2018). In addition, the diffusion of radicals and of the molecular products along the damage track vs. temperature and under irradiation is complex and not well understood (e.g., Benit & Brown 1990). It is also true that the release of $O_2$ trapped in ice is complex even in the absence of irradiation (e.g., Laufer et al. 2017), as it depends on the thermal-induced structural changes of the ice matrix. All of these issues affect our ability to model, for instance, how strongly bound, on average, an individual $O_2$ molecule is in the radiation-damaged ice. However, consistent with laboratory data, $O_2$ and $H_2O_2$ are observed as trapped species in the ice grains on the icy Jovian satellites (Spencer et al. 1995; Carlson et al. 1999a; Spencer & Calvin 2002; Hand and Brown 2013). Such observations are consistent with radiation damage studies in which under irradiation volatiles, such as $O_2$, can become trapped in radiation produced 'bubbles' (multiple vacancy sites/voids) in an ice grain (Johnson & Jesser 1997) or at defect sites (Loeffler et al. 2006) with the concomitant $H_2O_2$ being bound as more refractory molecules in an irradiated ice (e.g., Carlson et al. 1999a; Carlson et al. 2009). Although a large amount of important work on these topics still needs to be carried out, below we proceed based on our present, but incomplete, understanding.

**Observations**

By observing atomic oxygen emission lines using HST, Hall et al (1995; 1998) confirmed that Europa has an ambient gas consisting of both atomic and molecular oxygen. The column density of $O_2$ estimated from these observations was very roughly consistent with the amount produced as predicted by combining estimated plasma fluxes with the laboratory data for $O_2$ yields (Johnson et al. 1982). These early observations have been extensively expanded and are roughly consistent with the more recent observations (e.g., McGrath et al. 2009; Roth et al. 2016) as well as those of the New Horizon spacecraft (Retherford et al. 2007; see also Plainaki et al 2018 Fig. 4). The concomitant ambient $H_2$ may have also been observed indirectly (e.g., Mauk et al. 2003) as well as the dissociation of ejected gas phase $H_2O$ molecules resulting in Lyman alpha emission from the excited atomic hydrogen. The latter was the first evidence that outgassing of $H_2O$ might be occurring at Europa (Roth 2014) and has been used recently to describe Europa's hydrogen corona (Roth et al. 2017). Here we focus on $O_2$ observations.

The global source and loss rates of oxygen have also been modeled and discussed with the oxygen atmosphere exhibiting a near surface component dominated by $O_2$ and an extended component eventually dominated by O. Unfortunately the near surface plasma properties are still uncertain, affecting the calculation of both the radiolytic production of oxygen and its emission and loss rates. Therefore, estimates of the global $O_2$ source rate at Europa have varied due to the uncertainties of the nature of the particle flux as well as the actual physical conditions at Europa's surface as indicated in Fig. 1.

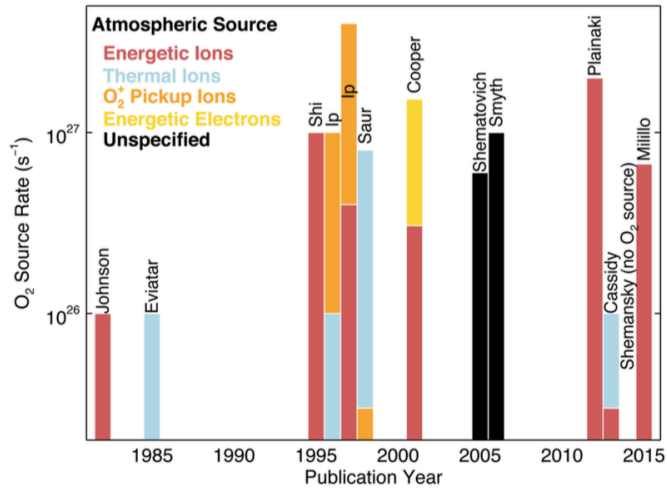

Figure 1. Estimates of global source O$_2$ source rates with the colors indicating the incident radiation type assumed (Cassidy 2016)

What is possibly more interesting, and not yet well understood, is the observed spatial morphology of the observed aurora, which is used to roughly extract the spatial distribution of the atmospheric molecular and atomic oxygen. Experiments indicate that the radiolytically produced O$_2$ leaves the surface with a speed distribution determined in part by its binding in ice and in part by the surface temperature (Johnson et al. 1983). For this reason it is not easy or maybe even possible to observationally distinguish between thermally desorbed and sputtered O$_2$. In addition, the distribution of the O$_2$ ejection speeds of O$_2$ is found to be much less than its escape speed. This results in an average hop distance of an ejected O$_2$ that is much smaller than Europa's radius. In this way the observed surface-bounded atmosphere should reflect, to some degree, the local surface and plasma properties. Since Europa's surface composition appears to vary with latitude and longitude, as does the incident plasma temperature and flux (e.g., Nordheim et al. 2018), it is unsurprising that the auroral data suggest that its oxygen atmosphere is non-uniform (Roth et al. 2016).

Variations in the surface reflectance spectra are suggested to be due to trace sulfur species (Carlson et al. 1999b; 2005) on the trailing hemisphere, whereas in the Polar Regions and leading hemisphere the ice appears to be cleaner and possibly dominated by trace carbon species (e.g., Carlson et al. 2009; Hand & Brown 2013). Indeed, observations suggest that there may be more atmospheric O$_2$ over the regions of cleaner ice (McGrath et al. 2004; Roth et al. 2016) but the observed north-south differences in the emission rate are suggested to be due primarily to the differences in electron temperature and density (e.g., Roth et al. 2016, Fig. 10). Because of the uncertainties in the effect of the spatial variations in the composition and the incident plasma, both of which affect the O$_2$ source rate, carrying out simulations of the atmosphere in order to obtain a detailed description of the emission observations has been problematic. This is not only due to the spatial distribution of radiolytic production and loss, but, as stated earlier, is also due to the uncertainty in the binding state of the O$_2$ formed in the ice, and its eventual fate after it is ejected and then returns to the regolith. Due to these complications, Oza et al (2018a) took a somewhat different approach to be discussed shortly.

The published HST/STIS images have a spatial resolution between ~71 km and ~95 km per pixel depending on Europa's distance, allowing one to separate emission features originating from the atmosphere within ∼500km from the surface and those feature originating from an extended corona as analyzed in great detail in Roth et al. (2016). The excitation rate depends, of course, on electron temperature, as seen nicely in Fig. 12 of Roth et al. (2016). For the expected electron temperatures near Europa, an intensity ratio of ~2 for the 1356A/1304A lines is indicative of electron impact dissociation of molecular oxygen. It was estimated that throughout its orbit 95-99% of the near-surface ($r < 1.25 r_E$) emission is due to electron impact of $O_2$, as opposed to electron impact of O (Roth et al. 2016). In addition to north-south asymmetries discussed, longitudinal asymmetries were consistently found at several orbital positions, $\phi_{orb}$, from the HST dataset spanning from 1999-2015 (see Fig.1a). The asymmetries shown were extracted by comparing *average* emission intensity over a range of latitudes and longitudes, but had no clear plasma or atmosphere explanation (e.g., Roth et al. 2016). Such asymmetries were difficult to discern from the patterns of individual images analyzed over the decades, but orbital and spatial averaging mitigates the uncertainties. One such orbital dependence described in Roth et al. (2016) was subsequently examined in Oza et al. (2018a, b): the near-surface dusk-over-dawn oxygen emission intensity ratio. It was used to test simulations of the *near-surface* $O_2$ column density.

Although there is considerable scatter in the data, it is seen in Fig. 2 that to first order, the dusk-over-dawn oxygen aurorae intensity ratios are *not* extremely sensitive to orbital position, $\phi_{orb}$, with a nearly independent average ~ 1.6, allowing for the considerable uncertainties. In this rough estimate, the 1356A line is used, as the effect of direct solar resonance scattering by the ambient O is negligible. That these observations do exhibit considerable scattering about the average is not surprising considering the spatial and temporal variability discussed above as well as the difficulty of observing near eclipse. In fact, the ratios near eclipse could be ignored. Below we consider the implications of this result. Before proceeding we note that, although the scatter in the data in Fig. 2 is large, especially near eclipse, the near-surface dusk/dawn emission ratio might be slightly larger in the vicinity of $\phi_{orb}$ ~180°. In this region of the orbit the dawn half-hemisphere is the darker, trailing hemisphere and the dusk half-hemisphere is the leading, brighter hemisphere. This would appear to be consistent with a somewhat enhanced ratio, possibly consistent with an enhanced source in the region of cleaner ice. In the following we ignore this possibility as more observations would be required to be definitive.

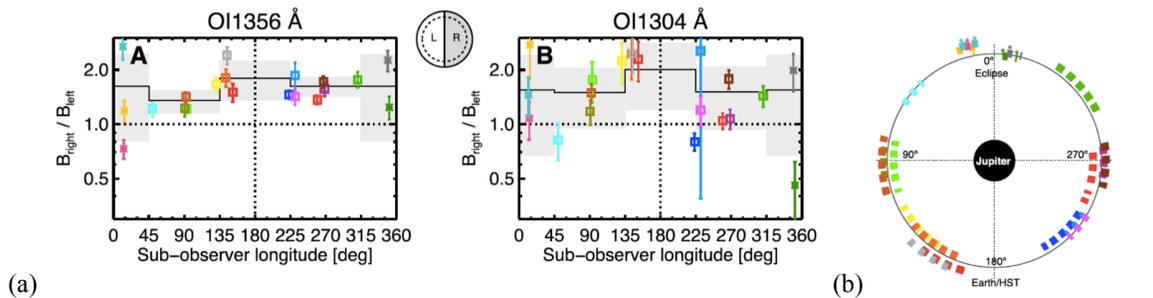

(a) (b)
Fig. 2: Data from Roth et al (2016) (a) Right/left (dusk/ dawn) half-hemisphere, near surface (∼<1.25$r_E$) brightness ratios in the two oxygen emissions lines indicated. The 1356A/1304A ratio at each sub-observer position, $\phi_{orb}$, varies but is near to 2 suggesting the emission is heavily dominated by the process $O_2 + e \rightarrow O^* + O$; (b) Color code for locations of Europa for data taken in 1999, 2012, 2014 & 2015.

**Simulations**

Assuming that Europa's atmosphere is marginally collisional or even collisionless, there have been a number of applications of the laboratory data in simulations of its atmosphere (e.g., Smyth & Marconi 2006; Shematovich et al. 2005; Saur et al 1998; Cassidy et al. 2013; Luchetti et al. 2016; Oza et al. 2018b). Laboratory data indicate there is a surface temperature and irradiation time dependence to the formation and ejection of $O_2$ by UV photons (e.g., Westley et al. 1995), by plasma electrons (e.g., Orlando & Kimmel 1997) and by plasma ions (e.g., Brown et al. 1982; Johnson et al. 1983; Teolis et al. 2017). Therefore, it was pointed out that the thermal and irradiation history must be accounted for in determining the radiolytic source rate (e.g., Teolis et al. 2005; 2009; Cassidy et al. 2013).

Unfortunately, it is difficult to carry out laboratory experiments in a sample that accurately represents a porous regolith (Cassidy and Johnson 2005). That is, the vapor deposited laboratory samples have a porosity at the submicron molecular scale, whereas the regoliths exhibit porosities at the 100s of micron scale. In addition, the charged particles used in many of the experiments had penetration depths smaller than the typical regolith grain size, whereas some experiments use incident particles with penetration depths greater than the typical grain size by irradiating relatively thin samples, in which case the substrate might play a role. Therefore, the concept of 'the surface' can differ in these experiments from that understood to be present on Europa or any other icy body. In addition, the thermal plasma electrons and ions at Europa primarily bombard the trailing hemisphere while energetic ions with energies above ~a few hundred keV bombard the surface much more uniformly (Cassidy et al., 2013). Energetic electrons on the other hand, bombard the surface in highly non-uniform patterns, leading to the formation of 'bombardment lenses' centered on the leading and trailing hemispheres at low latitudes (Paranicas et al. 2001; Nordheim et al. 2018) as indicated in Fig. 3. As energy deposition by energetic electrons dominates at cm to m depths, this degree of non-uniformity has important implications for the radiolytic production of $O_2$ at depth. Convolved with the significant variation in charged particle energy deposition vs. depth and surface location, is the variability of surface materials and trace species at different locations (e.g. as discussed, the leading hemisphere and polar regions appear to contain different trace species than the trailing hemisphere). For these reasons also agreement on the use of the laboratory data in the simulations remains problematic (see also Cassidy 2016; Plainaki et al. 2018).

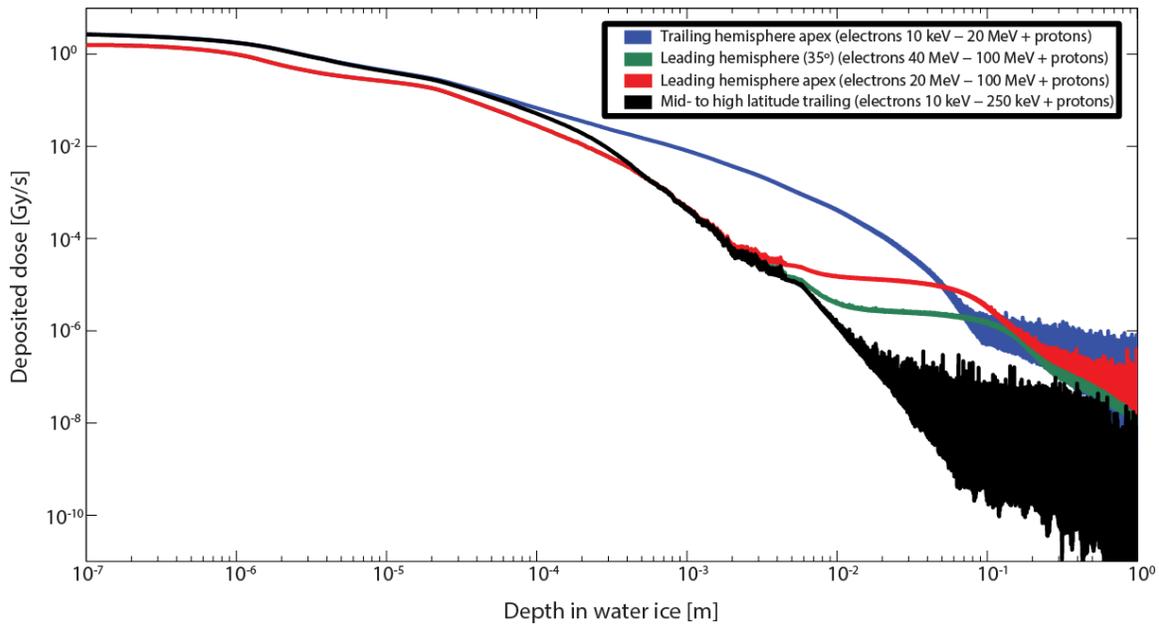

Fig. 3 Time to obtain a dose of 100eV/16amu vs. depth in m and vs. regions on Europa indicated in the inset. This dose can be used with the G-values (number of a product per 100eV deposited) published in a number of papers. (From Nordheim et al. 2018)

In spite of such uncertainties, there have been a number of efforts applying models of the laboratory data in collisionless Monte Carlo simulations of Europa's atmosphere. In a recent attempt at simulating the observed oxygen aurora, Oza et al. (2018b) tested nominal source distributions, including, for the first time, the effect of Europa's rotation about Jupiter. They also considered the effect of the varied spatial composition of Europa's surface on the sputtered $O_2$ as well as on its fate when it returned to the surface. In these simulations, the initially desorbed and thermally reemitted $O_2$ were tracked as they were ejected and ballistically hopped across the surface subject to gravity and the centripetal and Coriolis forces. Although the range of possible source processes and the range of the surface effects could surely be expanded beyond those used, the conclusion was that Europa's rotation about Jupiter was critical in generating the dusk-over-dawn asymmetry. However, the fact that the observed dusk/dawn hemispherical ratios did not vary strongly with orbital position, as seen in Fig. 2, could not be reproduced. That is, the models used gave near surface emission ratios that were strongly dependent on the orbital position, differing at some orbital positions from the HST ratios by an order of magnitude. Because the spatial dependence of the loss rate used in those simulations was a simple global average, it was also pointed out that one could carefully design a spatially variable source and a spatially variable loss rate that could certainly force the simulations to produce dusk/dawn ratios more uniform with orbital longitude as in Fig. 2. Since such a simulation would be artificial, a different approach was tried.

Although the physical processes are complex, the result in Fig. 2a based on the observations suggests some simple aspect was being overlooked. In order to focus on the role of Europa's orbital position with respect to the solar illumination, the surface rotation and the preferential bombardment of the trailing hemisphere, the longitudinal variation in the latitudinally averaged $O_2$ column density was modeled. This was done in order to directly extract the role of the longitudinal, rotational and thermal dependence of the likely source rate. Therefore, a simple rate equation, balancing source and loss rates, was used to estimate the dependence on the local longitude, $\phi$, of the time-variable, latitudinally averaged $O_2$ column density, $N(\phi, t)$, vs. Europa's orbital position, $\phi_{orb}$ (Oza et al. 2018a). As Europa rotates with angular speed, $\Omega$, phase-locked to Jupiter, $\phi_{orb}$ is simply related to the location of solar illumination angle, $\phi_s$, on Europa's surface. Because the average ballistic hop distance for an ejected $O_2$ is much smaller than Europa's radius and the lifetime of $O_2$ is shorter than the time for it to diffuse by hopping a significant distance across Europa's surface, a rate equation for the *local* $O_2$ column density, $N$, can be approximated as

$$dN(\phi,t)/dt \approx \Phi(\phi, \phi_s) - \nu(\phi, \phi_s)N(\phi,t) \qquad (1)$$

Here $\phi$ is the longitudinal position on the surface, $\Phi$ is the local $O_2$ flux from the surface and $\nu$ is the local loss rate. This ignores the hopping across the surface described in Oza et al. (2018b), which could be included by a diffusive term. However, in this form Eq. 1 is readily integrated over time as Europa rotates, allowing one to test the steady state spatial and thermal dependence of $O_2$ that is suggested by the spatial and thermal dependence of the source rate as well as on the loss rate. Although diffusion across the surface, as well as the centripetal and Coriolis forces, are ignored in Eq. 1, the suggested longitudinal and thermally dependent source and loss rates typically used gave near surface $O_2$ column densities that also depended very strongly on orbital position, as was seen to be the case in the detailed simulations in Oza et al. (2018b). This suggested that Eq. 1 could be used to test a number of assumed source processes. Using a spatially and thermally dependent *sputter source in* Eq.1 suggested by data from laboratory experiments, it is seen in Fig. 4 that the dusk/dawn ratio versus orbital position disagrees considerably with the ratios in Fig. 2 extracted from the emission observations. This was the case even allowing for the delays in emission suggested by laboratory measurements (Teolis et al. 2005; Cassidy et al. 2013).

However, $O_2$ trapped in an ice regolith can in principle be released thermally, as discussed frequently (e.g., Loeffler et al. 2006; Bar-Nun et al. 2013; Laufer et al. 2017), and as can also seen experimentally on warming an irradiated ice sample (Teolis et al. 2005). Therefore, a regolith permeated with $O_2$ that can be released thermally could in principal dominate the $O_2$ sputter source. Using an average loss rate, $\nu$, Oza et al. (2018a) noted that a source rate depending only on the solar illumination angle, $\phi_s$, could account for the weak dependence of the dawn/dusk emission ratio on the orbital position. That is, writing the surface desorption flux vs. $\phi_s$ in a clearly oversimplified form

$$\Phi(\phi_s) \sim \Phi_0[1 - \cos(\phi_s)] \qquad (2a)$$

where $\Phi_0$ is the maximum sublimation at noon, $\phi_s = \pi$ and goes to zero in eclipse, the steady-state, latitudinally-averaged column density only depends on $\phi_s$: $N(\phi) \sim (\Phi_0/2\nu)[1-(1+\beta^2)^{-1}(\cos(\phi_s) - \beta \sin(\phi_s))]$ with $\beta = \Omega/\nu$. Using this estimate of the column density the calculated half-hemisphere averaged ratio is found to be independent of $\phi_{orb}$. That is, on substituting Eq. 2a into Eq. 1, the half-hemisphere average of dusk/ dawn ratio, <R>, of the near surface, $O_2$ column densities is

$$<R> \sim [1+ (2/\pi)(\beta +1) + \beta^2] / [1 - (2/\pi)(\beta -1) + \beta^2]. \qquad (3)$$

Remarkably this depends only on β, the ratio of the rotation rate to the average loss rate, indicating that rotation is critical for the dusk/dawn asymmetry at Europa as shown in the simulations in Oza et al. (2018b). That rotation plays a critical role in the surface bounded atmospheres was noted much earlier for the lunar atmosphere (Hodges & Johnson 1968) and Mercury's Na atmosphere (Cassidy et al. 2015; Leblancand Johnson 2010). It is seen that in this approximation <R> is not only constant but is greater than one. That is, for an $O_2$ source that is determined primarily by the solar flux, the peak column is shifted from near noon toward dusk simply due to Europa's rotation.

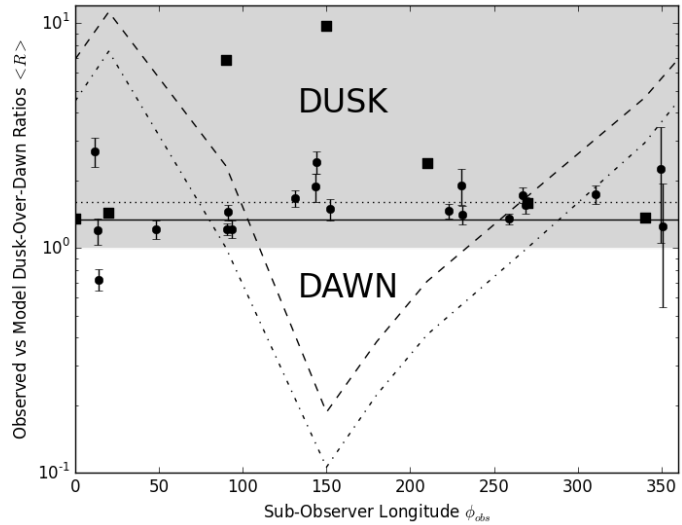

Fig. 4: Half-hemisphere averages of the dusk-over-dawn ratio of emission and modeled column densities <R> vs. sub-observer longitude. Circles: ratio of HST emission intensities from Roth et al. (2016) as in Fig. 2; average ~1.6. Squares: detailed simulations from Oza et al (2018b) in which the trailing hemisphere enhancement in the plasma flux (Cassidy et al 2013) onto the surface determines the source rate. Lines: column density ratios calculated with different source models and an average loss rate. Dotted-dashed & dashed lines: plasma ram sources with (dashed) and without (dotted-dashed) a temperature enhancement. Solid & dotted lines: purely thermal ratio <R> from Eq. 3 with $\beta = \Omega/\nu$ is 6.7 (solid) or 1 (dotted) giving <R> ~1.2 & 1.6 respectively (adapted from Oza et al. 2018a Fig. 3).

The scale height of the observed $O_2$ atmosphere has been discussed extensively and may be affected on the trailing hemisphere by collisions with sputtered water molecules (Oza et al. in prep). Yet it is agreed that the $O_2$ molecules have a scale height much smaller than Europa's radius. Therefore, it is not unreasonable that the averaged near surface $O_2$ column densities are roughly proportional to the averaged near surface component of the observed $O_2$ emission. Although that is a significant assumption, in Fig. 4 we compare the ratio in Eq. 3 to the observational ratios from Fig. 2. Using the loss rates based on photo and electron impact (Turc et al. 2014, Table 2) gives an ionization rate, $\nu \sim 3.1 \times 10^{-6}$/s for an electron density of 70 cm$^{-3}$. Using that value as a globally averaged loss rate and $\Omega = 2.1 \times 10^{-5}$/s at Europa, then $\beta \sim 6.7$. Substituting $\beta$ in Eq. 3 gives a half-hemisphere average dusk/dawn ratio of $<R> \sim 1.2$. Although $<R>$ in Eq. 3 is independent of $\phi_{orb}$, this underestimates the average ratio, $\sim 1.6$, obtained from the HST data. However, the sense of this result is consistent with an emission ratio nearly independent of orbital position. Recent modeling of the plasma flow (Dols et al. 2016) suggests that charge transfer with the ionized component of the atmosphere dominates the near surface loss rate of $O_2$. If that is the case, including all $O_2$ loss processes (e.g., Lucchetti et al. 2016), the net, globally-averaged loss rate can be as large as $\nu \sim 2.0 \times 10^{-5}$/s. This would result in $\beta \sim 1$ and $<R> \sim 1.6$. This is, in surprising, and possibly fortuitous, agreement with the *average* HST result in Fig. 2.

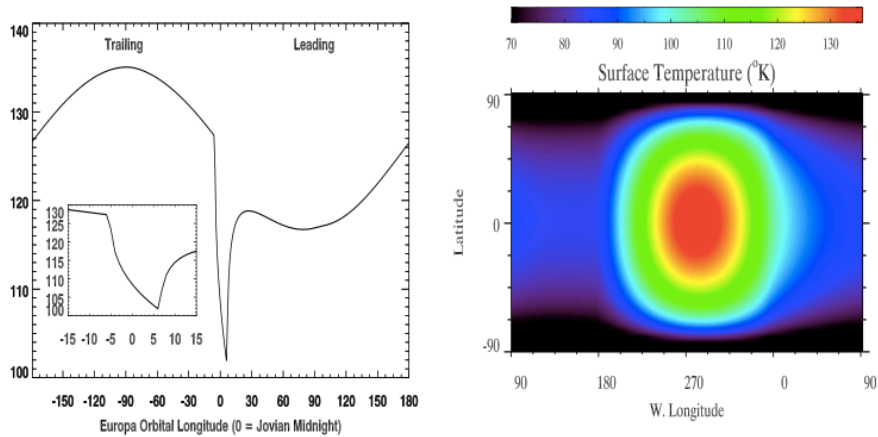

Fig. 5 (a) Orbital modulation of Europa's equatorial sub-observer temperature, due to its albedo and the Jovian shadow, in degrees K vs. orbital longitude (b) temperature map sub-observer longitude 270° vs. planetographic longitude and latitude at an obliquity roughly consistent with the Roth et al. observational data. The duration of the surface T drop at Jupiter's obliquity is small, but enough to matter. Depending on the season the eclipse duration is $\sim 3\%$ of a Europa day. The regolith T in this model has a skin depth $\sim 4$cm (Spencer 1987) and ignores a solid state greenhouse effect. (From Oza et al. 2018b, Fig. 2)

The source rate in Eq. 2 is, of course, overly simplified as there are leading and trailing differences in composition and albedo resulting in much more complicated temperature profiles as seen in Spencer et al. (1999) and in Fig. 5. Therefore, the form in Eq. 2 could be improved by using a sublimation rate that depended on the actual temperature. Assuming that the dominant source is indeed thermal desorption from a regolith permeated with adsorbed $O_2$, the source flux above, $\Phi(\phi_s)$, could be approximated, as is often the case, as proportional to $\exp(-U_s/kT)$ with $U_s$ an average activation/binding energy. Here we crudely approximate an equatorial surface temperature like that in Fig. 5a, averaged over a range of longitudes and latitudes due to thermal inertia, as $kT = kT_0 (1 - \varepsilon \cos \phi_s)$ with $\varepsilon = \Delta T/T_0$ and $\Delta T$, the temperature excursion. Assuming $\Delta T$ is small relative to $T_0$, the average equatorial temperature, we obtain a latitudinally-averaged source rate, $\Phi$, that is very roughly proportional to $[1 - (U_s/kT_0) \varepsilon \cos \phi_s]$. This is seen to have a dependence like that in expression in Eq. 2a, justifying its use to approximate a thermal desorption source. In instead of Eq. 2a, one could write,

$$\Phi(\phi_s) \sim \Phi_0[1 - c \cos(\phi_s + \Delta\phi_s)]/(1+c) \qquad (2b)$$

where c can be used to fit the difference in the noon/midnight ratio. In Eq. 2b, $\Delta\phi_s$ is included to account for the shift in the peak temperature from noon as seen in Fig. 5b. That is, due to the *thermal inertia* of the icy surface the peak in the surface temperature of Europa in the near equatorial region is seen to be shifted toward dusk. For a thermal source like that suggested here, the peak temperature is shifted towards dusk slightly increasing the averaged dusk/dawn ratio, $<R>$, which we estimate to be ~ 15%. In addition, a direct sputter source would add to any thermally-induced source. For instance, adding to Eq. 2a a global sputter source, $\Phi_s$, as suggested in Cassidy et al. (2013), $<R>$ in Eq. 3 becomes

$$<R> \sim [1 + \delta_s + (2/\pi)(\beta+1) + \beta^2]/[1 + \delta_s - (2/\pi)(\beta-1) + \beta^2] \qquad (4)$$

where $\delta_s = \Phi_s/\Phi_0$. Not surprisingly, this expression for the dusk/dawn ratio is also independent of orbital position, but is altered by $\delta_s$, the ratio of the direct sputter source to the peak-thermal source. Obviously, more realistic expressions for $\Phi(\phi_s)$ are needed, but only after we have a good description of the physics that is occurring. The above discussion simply shows that even with additional complexity a primarily thermally driven source is consistent with a critical aspect of the near-surface observations.

There are, of course, uncertainties in the average loss rate, $\nu$, as well as the sputter to thermal source ratio, $\delta_s$. Therefore, using the available observations to make a definitive statement on their relative importance is not likely at present. However, most previous simulations have assumed that $\delta_s$ is strongly dependent on orbital longitude. Therefore, a source rate that responds primarily to the local temperature as in Oza et al. (2018a) and not, primarily to an enhanced trailing hemisphere sputtering rate, appears to be a much simpler way to account for the lack of significant variation with Europa's orbital position of the averaged dusk/dawn emission ratio observed. Such a source would appear to suggest that Europa's regolith has become permeated with trapped $O_2$, which upon thermal desorption populates the near-surface atmosphere, with, of course, a smaller contribution from the ultimate plasma source which replaces and redistributes the $O_2$ lost by the escape, subduction, dissociation and ionization.

Clearly, additional improvements can be made by accounting for the differences in temperature and composition on the leading and trailing hemispheres as discussed. However, as seen from the data in Fig. 2a the fully illuminated trailing hemisphere (orbital longitude 270°) and the fully illuminated leading hemisphere (orbital longitude 90°) exhibit surprisingly similar radially-averaged dusk/dawn emission ratios, though with significant uncertainties. This is consistent with the fact that the difference in radial dependence of the emission intensity of the 1356A line at these orbital longitudes is not large, as seen in Fig. 5 (red lines: dashed leading and dotted trailing). Even the off-disk intensity averages in Fig. 6 are surprisingly similar for the 1304A line. The similar radial dependences for the illuminated leading and trailing hemispheres both of which increase with radial distance from the subsolar point is not inconsistent with a predominantly thermally driven source. The fact that the hemispherical differences seen in Fig. 6 for the 1304 A line are larger than those seen in the 1356A suggests that spatial variation in the plasma, affecting the dissociation of $O_2$, and the subsequent ionization and excitation of the dissociation product

O, does play a role. Therefore, corrections to the average loss rate, ν, in Oza et al. (2018a,b), and also used to obtain Eqs. 3 and 4 above should be accounted for. For this reason, a good description of the plasma temperature and particle flux *very close* to Europa's surface is needed. Although one could, in principle, reproduce the HST results in Figs. 4 and 6 by designing functions for the source and loss rates that depended on both $\phi$ and $\phi_s$, below we discuss the likelihood that a sublimation-dominated source is reasonable.

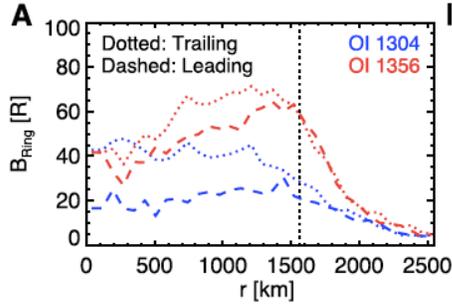

fig. 6: Radial brightness profiles of the mean OI 1356 Å (red) and OI 1304 Å (blue) brightness averaged over concentric 0.05 $R_E$ wide rings around the disk center (0 km) in all exposures: dashed, fully illuminated leading hemisphere, $\phi_{orb} \sim 90°$; dotted, fully illuminated trailing hemisphere, $\phi_{orb} \sim 270°$. From Roth et al (2016) Fig. 11a.

**Discussion**

As pointed out following the early laboratory measurements of the radiolysis of ice (Brown et al. 1982; Johnson et al. 1983), the preferential loss of the much more volatile $H_2$, implies that the irradiated ice becomes oxygen rich. At Europa this effect is enhanced by the fact that $H_2$, produced and desorbed during radiolysis, also readily escapes Europa's gravitational field, whereas the ejected $O_2$ primarily returns to the surface, as does a significant fraction of the O formed following dissociation of $O_2$ (e.g., Oza et al. 2018b). Therefore, it is clear that the regolith is oxygen rich, at least down to the penetration depth of the incident energetic particle radiation, ~ a meter (e.g., Cooper et al. 2001; Paranicas et al. 2003; Norheim et al. 2018). Because the regolith is porous, diffusion would suggest that over long time periods the ice could be oxygen rich to much larger depths (e.g., Johnson et al. 2003).

Although the state of the oxygen rich ice is not well defined, and likely differs between hemispheres and with depth, $O_2$ *has been* observed as trapped in Europa's surface ice grains (Spencer and Calvin 2002). That observation was associated with a band from a solid-like γ phase $O_2$ at 5773Å. Because solid $O_2$ cannot be sustained in vapor pressure equilibrium at Europa's surface temperatures (e.g., Fray and Schmitt 2009), it was suggested to be trapped in radiation produced voids (Johnson & Jesser 1997). Recently, that band was observed to exhibit an unexplained spatial and temporal variability at Europa as shown (Spencer & Grundy 2017). This is likely related to a *combination* of plasma irradiation and local temperature affecting trapping (e.g., Shi et al. 2009) and, possibly, water vapor deposition from the plumes.

The size of the grains, which appears to decrease by more than a factor of 10 from the trailing and to the leading hemisphere (see Cassidy et al. 2013, Fig. 10), can influence the $O_2/H_2O$ ratio inferred from the 5773Å $O_2$ observations. Assuming the band depth is due to $O_2$-γ, based on the expression derived by Hand et al. (2006) and the data in Spencer and Grundy (2016), then the trapped $O_2/H_2O$ ratio seems, quite surprisingly, to be within a factor of a few of that emitted from Comet 67P/ Churyumov–Gerasimenko: $O_2/H_2O$ (Bieler et al. 2015) and Comet 1P/Halley (Rubin et al. 2015). In addition to the $O_2$ observed at Europa in the 5773Å band, individual $O_2$ can be trapped at defect and dangling bond sites produced by the irradiation (e.g., Cassidy & Johnson 2005; Loeffler et al. 2006; Johnson 2011; Johnson et al. 2013) including at grain surfaces (Teolis et al. 2005). Indeed laboratory experiments indicate that the surface properties play a role in controlling the production and subsequent ejection of $O_2$ (e.g. Teolis et al. 2006; Teolis et al. 2017). The $O_2$ is seen to trap in the near surface region of the irradiated laboratory ice samples, which on Europa would be the surface of an ice grain exposed to the incident plasma. This $O_2$ is driven into the gas phase with the amount and rate depending on the irradiation history (Teolis et al. 2005). Indeed on irradiation followed by warming, $O_2$ is also seen to evolve (e.g., Teolis et al. 2009; 2017) and warming reduces dangling bond sites (Loeffler et al. 2006). Therefore, the production and release of $O_2$ is clearly temperature dependent, although for the ~135K maximum temperatures expected at Europa, outgassing in laboratory samples in which $H_2O$ and $O_2$ are co-deposited was estimated to be small (Teolis et al. 2005). Emission of $O_2$ from non-irradiated samples of co-deposited $O_2$ and $H_2O$ has been related to the crystalline phase transitions (e.g., Laufer et al. 2017), which occurs due to the rearrangement of hydrogen bonds.

Although the formation and ejection of $O_2$ from a laboratory water ice sample due to the incident radiation is initially negligible, it grows with irradiation time (e.g., Brown et al. 1982; Wesley et al. 1995; Teolis et al. 2017). This is unlike what occurs in gas phase $H_2O$ and is due to the ability of the solid to accumulate trapped precursors (e.g., Johnson 2011). Therefore, a significant amount of solid-state chemistry occurs before the $O_2$ yield reaches steady state and precursor trapping has been shown to be critical in modeling of the measured thermal and fluence dependences using *chemical rate equations* (e.g., Johnson et al. 2005; Johnson 2011). Since the nature of the trapping of both the chemical precursors and the $O_2$ is critical, it depends on the availability of defects/dangling H bonds, and has been shown, not surprisingly, to be enhanced by irradiation (Loeffler et al. 2006).

**$O_2$ Desorption Energy**

The laboratory measurements of the steady state yield of $O_2$ exhibit a roughly exponential temperature dependence with an apparent activation energy ~0.06eV found in a number of experiments (Reiman et al. 1984; Brown et al. 1984; Fama et al. 2008; Teolis et al. 2017). This activation energy, smaller than the typical hydrogen bond energy in ice and even smaller than the sublimation energy for solid $O_2$, is not yet explained. Since it is extracted from irradiation measurements, it is likely related to *radiation-enhanced* activation and/or diffusion.

Comparing the convenient, but overly simplified, model in Eq. 2a to the form for $\Phi(\phi_s)$ discussed prior to Eq. 2b, would imply that $(U_s/kT_0) \varepsilon \sim 1$ where $\varepsilon = \Delta T/T_0$. Using $\Delta T \sim 10K$ and $T_0 \sim 125K$ estimated from Fig. 5, a very approximate activation/bond energy can be obtained, $U_s \sim 0.14eV$. This is similar to hydrogen bond energies in ice, which, of course, vary somewhat depending on the ice structure. This is roughly 1.5 times as large as the sublimation energy for pure $O_2$. In the temperature range of interest on Europa, the outgassing from the co-deposited $H_2O/O_2$ was slow, as discussed, until the temperature was increased to the phase transition ~ 155K. However, over the relevant temperature range for Europa the co-deposited samples were shown to steadily release $O_2$ (Fig. 4 in Laufer et al 2017). A desorption energy, possibly fortuitously, close to 0.14 eV was extracted from co-deposition experiments in support of the ROSINA measurements (e.g., Laufer et al. 2017; Table I). These crude comparisons suggest that thermal desorption of $O_2$ attached to dangling H bonds might be the dominant source of the observed near surface $O_2$ at Europa.

The early measurements of the sputtered $O_2$ speed distribution (Johnson et al. 1983; Reimann et al. 1984) indicted that most of the ejected $O_2$ returned to Europa's surface as mentioned earlier. Therefore, in their extensive simulations Oza et al. (2018b) were concerned with the re-condensing $O_2$. They showed the heat of adsorption on the re-condensing $O_2$ molecules significantly affected their rate of desorption back *into* Europa's atmosphere. Because of surface compositional differences, the effective binding can vary, affecting the spatial morphology of local $O_2$ column density (e.g., Fig. 4 in Oza et al. 2018b). Therefore, new simulations need to be carried out based on the thermal desorption concept suggested in Oza et al. (2018a) in which one treats the nature of the trapping and the temperature vs. depth and time more carefully. That is, based on the above discussion, the local source is likely dominated by thermal desorption of $O_2$ trapped primarily at dangling hydrogen bonds from a regolith permeated with $O_2$ and with a smaller contribution from direct sputtering.

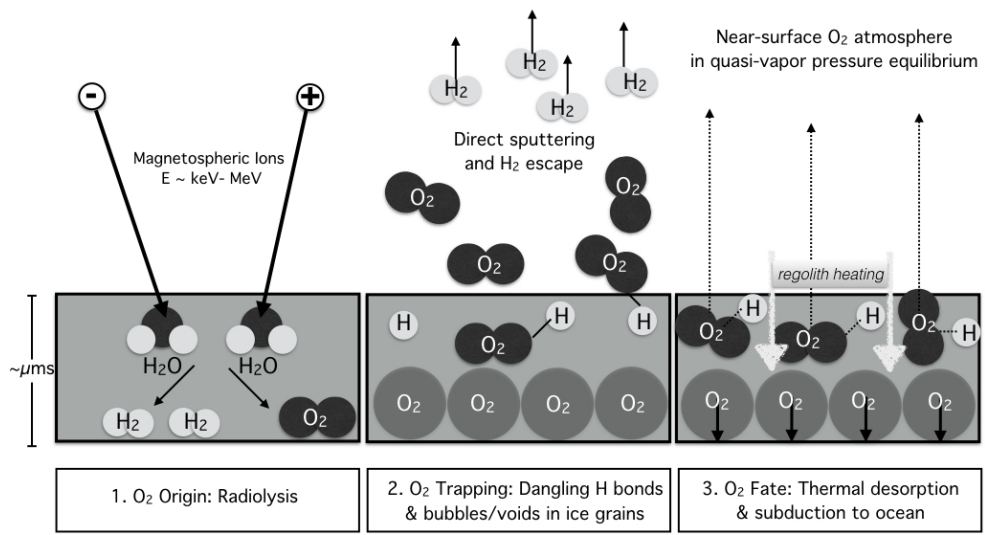

Figure 7: Schematic diagram of $O_2$ trapping and thermal desorption: 1) Primary origin of $O_2$ (and $H_2$) is magnetospheric ion radiolysis. 2) Due to preferential loss of $H_2$, the regolith becomes oxygen rich enhancing the production of $O_2$. Formed and returning O2 can become trapped at incomplete (dangling) H bonds (shown) as well as in voids (as shown and observed by Spencer & Calvin 2002). 3) The accumulated $O_2$ can then be thermally desorbed from the weak dangling bonds due to solar heating, maintaining a quasi-vapor pressure equilibrium (Oza et al. 2018a), with a smaller gas-phase contribution from direct sputtering of $O_2$. A fraction of the trapped $O_2$ is likely subducted.

**Summary**

As pointed out, the state of the trapped $O_2$ in Europa's regolith, not only affects the observation of the gas-phase $O_2$, but also affects its ability to diffuse downward and permeate the regolith. This is determined by the temperature variation with depth. That is, the local temperature gradient determines how any produced or returning $O_2$ diffuses through the ice matrix via defect or grain boundary diffusion (Johnson & Quickenden 1997; Johnson 2011). The presence of oxygen rich material at depth is also affected by the radiation which modifies the, top ~ meter of Europa's ice mantle as discussed above. This radiation not only produces an oxygen rich ice, but also can affect the crystal structure producing dangling bonds and causes radiation-enhanced diffusion. Since the porosity likely decreases with depth (e.g., Pappalardo et al. 2009), it is critical that the modeling needs to be carried out on the bulk processes and not just the near surface processes.

Although direct diffusion to the depth of the ocean is likely problematic, geologic mixing and subduction of oxygen rich ice has been suggested as a possible source of oxidants for putative ocean biology (Johnson et al. 2003; Grieves & Orlando 2005; Greenberg 2008; Russell et al. 2017). These are also processes which might bring any ocean organics to the surface where they might be detected (Johnson and Sundqvist 2008). Any oxidants transported to the ocean would not only be $O_2$, but other related and stable radiation-induced oxidants such as $H_2O_2$, sulfates, carbonates, carbonic acids, etc. which can yield an ocean with high chemical potential (e.g., Hand et al. 2006; Vance et al. 2016).

The simple paradigm for production of the observed $O_2$ atmosphere described here needs to be improved upon to assist in the planning of the observations of the JUICE and Europa Clipper missions. However, the concept considered suggests a fact that should not be surprising. That is, due to the considerable radiolytic processing of Europa's surface, and Europa's significant gravity, it is not surprising that Europa's ice mantle is permeated with oxidants and trapped $O_2$. This was, of course, the prime motivation for the suggestion that radiolytic processing of the surface ice combined with geologic activity might lead to subduction of oxidants to Europa's ocean (Chyba 2000; Johnson et al. 2003; Greenberg 2008; Russell et al. 2007)..

If Europa's regolith is permeated with trapped $O_2$, this might at first appear inconsistent with the Cassini observations at Dione and Rhea. Teolis et al. (2016) showed that the effective source of $O_2$ was *much smaller* than expected based on combining laboratory data with Cassini spacecraft measurements of the plasma flux (Johnson et al. 2008). However, the plasma fluxes at these objects are orders of magnitude smaller than at Europa, these bodies are on average far colder, and the surfaces appear to be older indicating less geologic activity than observed, for instance, at Enceladus (Porco et al 2006) and Europa (Roth et al. 2014; Sparks et al. 2016; 2017). This suggests that the application of the interesting and useful laboratory data on the gas-phase production of species produced by radiolysis must be applied carefully at each body, accounting for the local environment and the surface properties ----- a not very surprising conclusion. Therefore, we suggest new simulations need to be carried out based on the premise that Europa's near surface regolith, consisting of ice grains sizes that vary with depth, and with near surface thermal processing that varies with time and depth, is permeated with trapped $O_2$ that is primarily formed by the incident Jovian plasma. That is, molecules ejected by sputtering primarily return to the surface where they can diffuse to binding sites of varying strength and react with trace species. Subsequent radiation can drive the trapped $O_2$ back into the gas phase, but, as suggested in Oza et al. (2018a), thermal desorption of weakly trapped $O_2$, likely at dangling H bonds, appears to dominate the observed, near surface $O_2$ atmosphere.

**Acknowledgements**

The work at the University of Virginia was supported by a grant from NASA PDS program.